\documentclass[%
 reprint,
 amsmath,amssymb,
 aps,
]{revtex4-2}

\usepackage{graphicx}
\usepackage{dcolumn}
\usepackage{bm}
\usepackage{xcolor}

\begin{document}

\preprint{APS/123-QED}

\title{Measurement of the $^{99}$Tc $\beta$ spectrum with Silicon Drift Detectors} 

\author{Andrea Nava$^{1,2}$}
 \email{andrea.nava@mib.infn.it} 
\author{Andrea Del Contrasto$^{1}$}
\author{Leonardo Bernardini$^{3,4}$}
\author{Matteo Biassoni$^{2}$}
\author{Tommaso Bradanini$^{1,2}$}
\author{Chiara Brofferio$^{1,2}$}
\author{Marco Carminati$^{3,4}$}
\author{Silvia Capelli$^{1,2}$}
\author{Francesco Cappuzzello$^{7,8}$}
\author{Manuela Cavallaro$^{8}$}
\author{Massimiliano Clemenza$^{2}$}
\author{Giovanni De Gregorio$^{5,6}$}
\author{Giulio Gagliardi$^{1,2}$}
\author{Giorgio Grosso$^{9}$}
\author{Nunzio Itaco$^{5,6}$}
\author{Irene Nutini$^{2}$}
\author{Alessia Pozzi$^{1,2}$}
\author{Nicola Zorzi$^{10}$}
\affiliation{%
 \mbox{$^1$Dipartimento di Fisica G.Occhialini, Università di Milano-Bicocca, Milano, 20126, Italy} \\
 \mbox{$^2$INFN, Sezione di Milano - Bicocca, Milano, 20126, Italy} \\
 \mbox{$^3$DEIB, Politecnico di Milano, Milano, 20133, Italy}\\
 \mbox{$^4$INFN, Sezione di Milano, Milano, 20133,Italy}\\
 \mbox{$^5$Dipartimento di Matematica e Fisica, Università degli Studi della Campania “Luigi Vanvitelli”, Caserta, 81100, Italy}\\
 \mbox{$^6$INFN, Sezione di Napoli, Napoli, 80126, Italy}\\
 \mbox{$^7$Dipartimento di Fisica e Astronomia “Ettore Majorana”, Università di
Catania, Catania, 95123, Italy}
\mbox{$^8$INFN – Laboratori Nazionali del Sud (LNS), Catania, 95123, Italy.}
\mbox{$^9$LENA – Laboratorio Energia Nucleare Applicata, Università di Pavia, 27100, Pavia, Italy}
\mbox{$^{10}$Fondazione Bruno Kessler (FBK), 38123, Trento, Italy}
}%

\date{\today}

\begin{abstract}
The need for reliable calculations of Nuclear Matrix Elements is compelling for the next generation of neutrinoless double-beta decay experiments. This requires nuclear models to be validated against experimental data, such as non-unique forbidden $\beta$ decays, which have been found sensitive to details in nuclear calculations, most importantly to the renormalization of the axial and vector currents. 
We report here a measurement of the 2$^{nd}$ forbidden $^{99}$Tc $\beta$ spectrum performed for the first time with Silicon Drift Detectors, state-of-the-art semiconductor detectors for low-energy spectroscopy. We designed a novel hybrid spectrometer using a LYSO crystal read by a SiPM to precisely calibrate our main detector and to accurately measure the background. We then compared our measured spectrum with one obtained using cryogenic calorimeters, as well as with predictions from the Realistic Shell Model. 
Starting from Realistic Shell Model calculations performed with Bare decay operators, we carried out a Bayesian analysis to extract the average quenching factors required to reproduce both the measured spectral shape and the experimental half-life, obtaining $q_{g_A}=0.40(1)$ and $q_{g_V}=0.47(1)$. These values quantify the average renormalization of the axial and vector currents, respectively, and were compared with those predicted by RSM calculations employing Effective decay operators, thereby providing a benchmark for assessing the ability of the model to describe the second-forbidden $\beta$ decay of $^{99}\mathrm{Tc}$. More broadly, this comparison tests the reliability of the theoretical framework also used to predict $0\nu\beta\beta$ nuclear matrix elements.

\end{abstract}

\maketitle


\section{\label{sec:intro} Introduction}
Neutrinoless double-beta decay ($0\nu\beta\beta$) represents one of the most sensitive probes of neutrino nature and, more in general, of the physics beyond the Standard Model \cite{Avignone2008,DellOro2016,Henning2016,RevModPhys.95.025002}. However, the design of experiments searching for this decay depends on reliable nuclear matrix elements for the candidate isotopes. Since these cannot be measured directly, they must be calculated using nuclear-structure models, whose uncertainties affect sensitivity estimates \cite{RevModPhys.95.025002}. Experimental observables capable of benchmarking the underlying nuclear models and their treatment of weak-current operators are therefore essential.  
In this context, the precise characterization of $\beta$-decay spectra plays a central role \cite{Barea2013, Hayes2014, hyj7-l22h}, with non-unique forbidden $\beta$ decays providing particularly sensitive probes \cite{Haaranen2016,Kostensalo2017,Kostensalo2021, Kostensalo2023,DeGregorio2024}. Indeed, their half-lives and spectral shapes are highly sensitive to nuclear wave functions and to the effective renormalization of weak-current operators. Moreover, unlike allowed Fermi and Gamow–Teller transitions, these decays probe higher multipoles and interference between vector and axial-vector currents. Taken together, these features make forbidden $\beta$ decays stringent tests of the nuclear models employed in $0\nu\beta\beta$ calculations.

For these reasons, over the past decade, increasing attention has been devoted to the so-called spectrum-shape method, which exploits high-precision measurements of $\beta$ spectra to extract information on  
the renormalization of the axial and vector currents,
and benchmark theoretical predictions \cite{PhysRevC.95.024327, Kostensalo2021}. Despite significant progress, achieving a consistent description of both spectral shapes and decay half-lives across different isotopes remains a major challenge. \\
Several experimental studies have therefore investigated $\beta$-decay spectra across a range of isotopes and degrees of forbiddenness \cite{Dawson2009Cd113,BodensteinDresler2020,Kostensalo2021,Leder2022,Pagnanini2023,Paulsen2024,Belli2026Rb87}. Different experimental approaches have been developed for these measurements. Among these, cryogenic calorimeters provide excellent energy resolution and allow the radioactive source to be embedded directly within the detector \cite{Loidl2019}, enabling precise measurements of both spectral shapes and half-lives \cite{PhysRevLett.133.122501}. However, their operation at millikelvin temperatures introduces significant technical complexity and limits flexibility in source selection. Moreover, the analysis of the measured spectra requires a background model, often obtained through Monte Carlo simulations.

\begin{figure}[h]%
\centering
\includegraphics[width=0.5\textwidth]{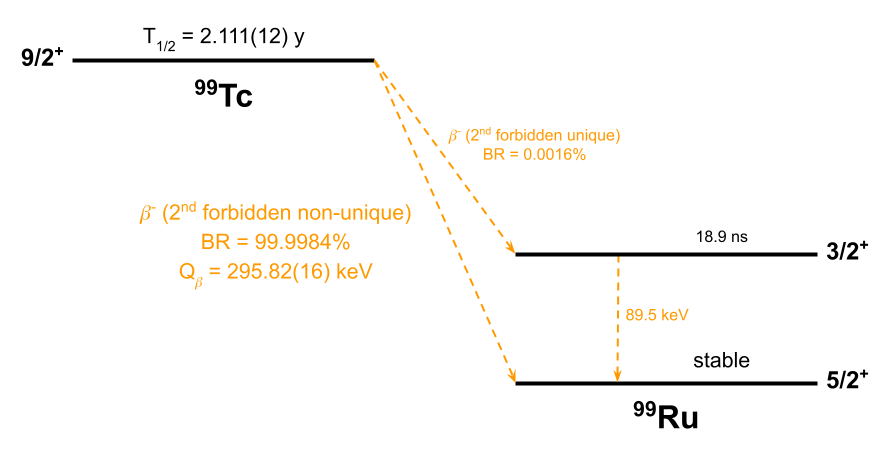}
\caption{Scheme of the $^{99}$Tc decay.}
\label{scheme}
\end{figure}

Within this broader experimental effort, the second-forbidden non-unique $\beta$ decay of $^{99}$Tc represents a particularly interesting case \cite{Paulsen2024,song2025measurement99tcbetadecayspectrum}. This interest is also motivated by the proximity of $^{99}$Tc to $^{100}$Mo, one of the leading candidate isotopes for $0\nu\beta\beta$ searches and the target of several experiments, including NEMO-3, AMoRE, CUPID-Mo, and the future ton-scale CUPID experiment \cite{Alenkov2019AMoRE,Bhang2012AMoRE,Arnold2007NEMO3,Armengaud2020CUPIDMo,Armengaud2021CUPIDMo,CUPID:2025avs}. Indeed, in parallel, substantial theoretical effort has been devoted to calculating the $^{100}\mathrm{Mo}$ nuclear matrix element within different nuclear-structure frameworks \cite{Rodin2003,PhysRevLett.105.252503,Barea2013,Hyvarinen2015,Wang2021,Coraggio2022}.  The $^{99}\mathrm{Tc}$ spectrum therefore provides a valuable experimental benchmark for assessing nuclear models employed in these calculations.

A scheme of the $^{99}$Tc $\beta$ decay is shown in Figure \ref{scheme}. The $\beta$ spectrum of $^{99}$Tc was measured by the MetroBeta collaboration using state-of-the-art thermal detectors in terms of energy resolution, known as Magnetic Metallic Calorimeters (MMCs) \cite{Paulsen2024}. Analysis of the MMC spectrum yielded an effective value of $g_A$ ($g_V$) of 0.574(36) (0.376(5)). 

A further independent measurement of the $^{99}$Tc $\beta$ spectrum using MMCs has recently been reported by the gA-EXPERT collaboration,  although the corresponding study has not yet been published \cite{song2025measurement99tcbetadecayspectrum}.

We present here a complementary approach based on Silicon Drift Detectors (SDDs), state-of-the-art semiconductor detectors for low-energy spectroscopy. SDDs offer an alternative route, combining good energy resolution with room-temperature operation and higher experimental versatility, allowing the study of a wide range of isotopes, as well as a direct measurement of the background. Silicon Drift Detectors (SDD) have been demonstrated to be good candidates for $\beta$-spectroscopy in the context of the TRISTAN \cite{Siegmann_2024} and ASPECT-BET projects \cite{Nava:2024wsa}. Nevertheless, such measurements introduce additional challenges related to electron interactions in the source and detector materials, including autoabsorption, backscattering, and incomplete charge collection, all of which can significantly distort the observed spectral shape. \\
A precise understanding of these effects is therefore essential to relate the measured spectrum to the underlying $\beta$-decay spectrum and to compare it meaningfully with theoretical predictions. This requires detailed modeling of both the detector response and the experimental geometry, which can be achieved using Monte Carlo simulations optimized for low-energy electromagnetic interactions. \\
In this work, we report the first measurement of the  $^{99}$Tc $\beta$-decay spectrum performed with SDDs. The analysis combines an accurate calibration of the energy scale, a direct measurement of the background, and a detailed GEANT4-based description of the detector response and source-related effects. 
The measured spectrum is analyzed through a Bayesian fit based on Realistic Shell-Model (RSM) calculations performed with bare decay operators, treating $g_A$ and $g_V$ as free parameters. The relativistic vector nuclear matrix element is fixed according to the CVC hypothesis \cite{BehrensBuhring1971}, as done in Ref. \cite{Paulsen2024}. The best-fit values of $g_A$ and $g_V$ determine the average quenching factors of the axial and vector currents, respectively, required to reproduce both the measured spectral shape and the $^{99}\mathrm{Tc}$ half-life. These experimentally inferred values are compared with the average renormalization predicted by RSM calculations employing effective decay operators derived microscopically and consistently with the effective Hamiltonian \cite{Paulsen2024}. 



\section{\label{sec:setup} Experimental setup}
\begin{figure}[h]%
\centering
\includegraphics[width=0.3\textwidth]{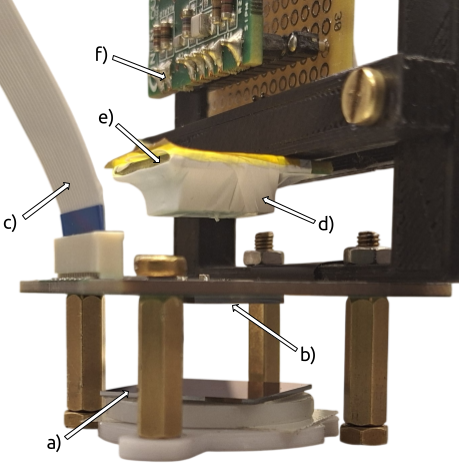}
\caption{Experimental setup used for the $^{99}$Tc measurement. In the picture, the $^{99}$Tc source (\textbf{a}), the SDD (\textbf{b}), the SDD bias and signal cable (\textbf{c}), the LYSO crystal used for calibration (\textbf{d}), the SiPM (\textbf{e}), and its electronic board (\textbf{f}) are visible.}
\label{setup}
\end{figure}

\textit{Source preparation} - The radioactive source has been prepared by dropping NH$_4$TcO$_4$, containing the isotope of interest $^{99}$Tc, on a 0.5 mm thick silicon substrate. The drops have been subsequently frozen by putting the sample in contact with a cold plate at -10 °C, and dried in a vacuum chamber. This technique is known as freeze-drying \cite{BRANGER2008685}, and it has been employed to ensure the spatial homogeneity of the source and to prevent the formation of large crystals, which would be problematic due to the large autoabsorption. The produced $^{99}$Tc source, mounted in the experimental setup, is indicated with \textbf{a} in Figure \ref{setup}. The source activity is $\sim$ 10 Bq. \\

\textit{Detectors} - This measurement was taken using a detection system specifically developed for beta spectroscopy. The SDD was fabricated by FBK as a square of 8 mm side and thickness of 1 mm. The large thickness is beneficial for a silicon detector for beta spectroscopy as it fully contains electrons up to several hundred keV. The large area is needed to minimise the number of events affected by incorrect energy reconstruction due to beta electrons interacting with the detector border. The shape of the electric field in this region can, indeed, cause charge loss. Furthermore, SDDs have a very thin entrance window of around 10 nm \cite{GUGIATTI2020164474}, which reduces the energy lost by beta electrons in this layer to an almost negligible level. These characteristics make SDDs suitable for beta spectroscopy at energies of O(100) keV. The SDD used is visible in Figure \ref{setup}, indicated by \textbf{b}, and facing the $^{99}$Tc source. The cable providing the SDD bias and connecting the front-end integrated amplifier CUBE, wire-bonded to the SDD, to the external biasing and conditioning board is also shown, indicated by \textbf{c}. \\
To make precise measurements of beta spectra, the energy scale of the SDD must be accurately known. Most precise beta spectrum measurements in the literature use radioactive X-ray or gamma-ray sources to calibrate the detector. To avoid biases arising from gain drifts, the calibration sources are typically measured alongside the beta spectrum. This is the case for the MetroBeta and gA-EXPERT measurements. However, to properly analyse the beta spectrum, the contribution of the calibration source must then be subtracted. This procedure is often realised through Monte Carlo simulations, thereby introducing a new systematic effect. To address this issue, we used a LYSO crystal containing the isotope $^{176}$Lu as a calibration source. The decay of $^{176}$Lu  provides several well-defined photon lines that can be used to calibrate the SDD. In particular, the decay can populate excited states of $^{176}$Hf, which subsequently de-excite through $\gamma$-ray emission. Additional X-rays can also be produced by atomic relaxation in the daughter atom or in other atoms of the crystal.  The photon energies used for the SDD calibration are listed in Table \ref{LYSO}. 

\begin{table}[h]
\renewcommand{\arraystretch}{1.3} 
\begin{tabular}{cc}
\hline
Energy (keV) & Origin \\
\hline
55.786 & Hf X-ray \\
201.83 & $^{176}$Hf $\gamma$-ray \\
306.78 & $^{176}$Hf $\gamma$-ray \\
\hline
\end{tabular}
\caption{Energies of the photons from $^{176}$Lu decay used for SDD calibration.}
\label{LYSO}
\end{table}

The LYSO crystal is well suited for this purpose, as it is a scintillator producing approximately 30000 photons per MeV. In our setup, the crystal was coupled to a silicon photomultiplier (SiPM) using optical grease, and both were wrapped in Teflon to improve light collection. The LYSO crystal, the SiPM and the SiPM in-vacuum electronics can be seen in Figure \ref{setup}, indicated respectively with \textbf{d}, \textbf{e}, \textbf{f}. 
After the beta decay of  $^{176}$Lu, a scintillation signal produced by the beta electron is always detected by the SiPM. Therefore, it is possible to isolate events from X-rays and gamma rays in the SDD by requiring a coincidence with the SiPM. Conversely, events in the SDD that are not coincident with the SiPM are considered unrelated to the calibration source, and mostly originate from the forbidden beta decay of $^{99}$Tc. This technique enables us to obtain a clean spectrum for calibration and a clean $^{99}$Tc spectrum for spectral analysis. Since calibration data are acquired together with $^{99}$Tc data, any corrections due to gain drifts during the measurement can be computed from the calibration data and applied directly to the $\beta$ spectrum. \\
Both the SDD and the SiPM were operated in a vacuum chamber at a pressure of $\sim$$10^{-5}$ mbar, to prevent electrons' energy loss in air before interacting with the detector, and the SDD was cooled to -10 °C to reduce leakage current, a procedure beneficial for energy resolution and dead time induced by pre-amplifier reset.

\section{\label{sec:data} Data taking and processing}
\begin{figure}[h]%
\centering
\includegraphics[width=0.5\textwidth]{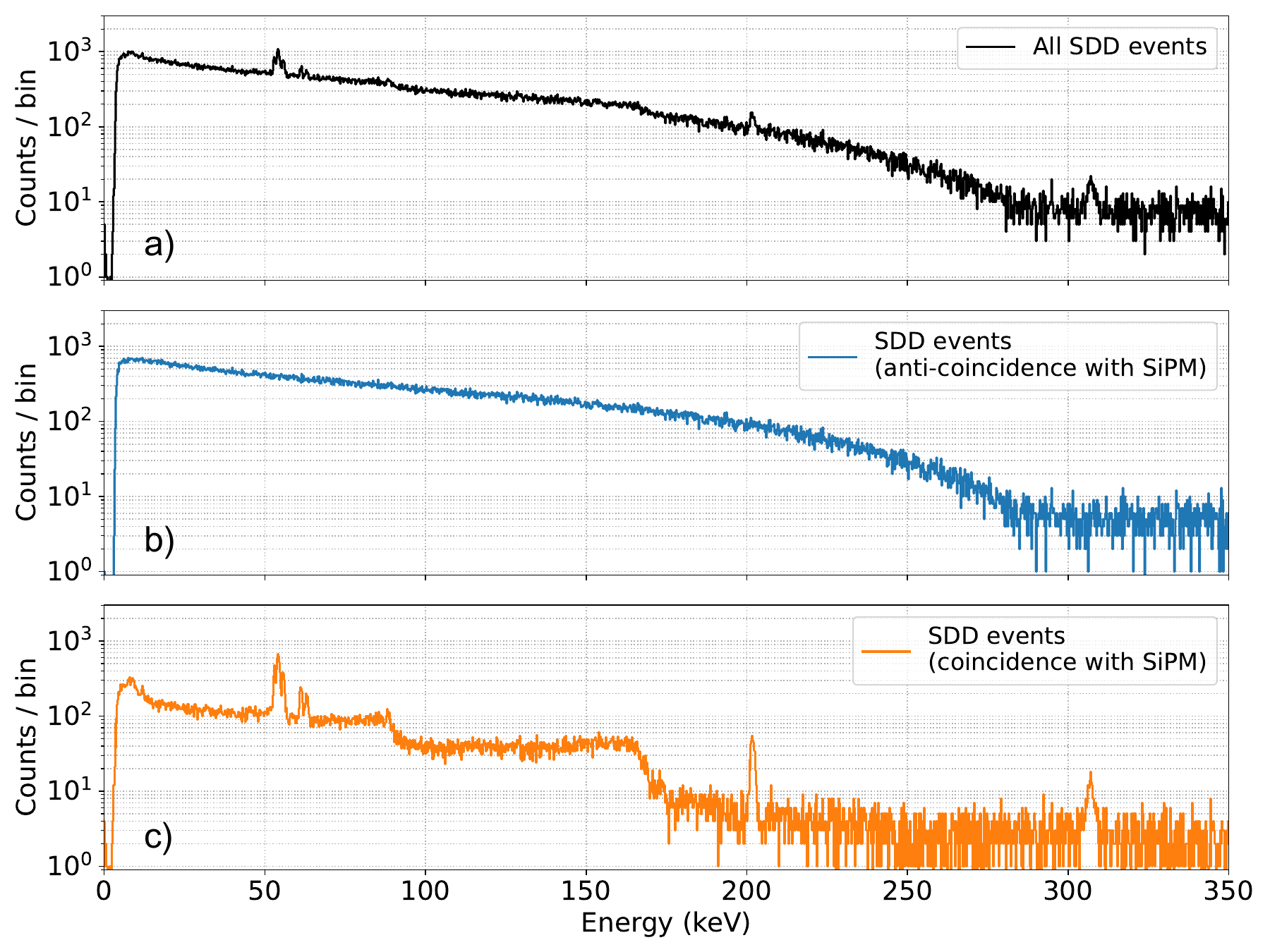}
\caption{Energy spectra measured with the SDD. a) All the measured events. b) Events in anti-coincidence with the SiPM (mainly $^{99}$Tc beta electrons). c) events in coincidence with the SiPM (mainly X and $\gamma$ rays from $^{176}$Lu decay).}
\label{spectra}
\end{figure}

The signals from the SDD and the SiPM were digitised at 125 MHz using a CAEN VX2740 digitizer and processed with a trapezoidal filter to extract the pulse amplitude. Additional information, such as the rise time of SDD pulses and a coincidence flag between the two detectors, was also computed. \\
We performed a six-day measurement, and the resulting spectra are shown in Figure \ref{spectra}. 

\begin{figure}[h]%
\centering
\includegraphics[width=0.5\textwidth]{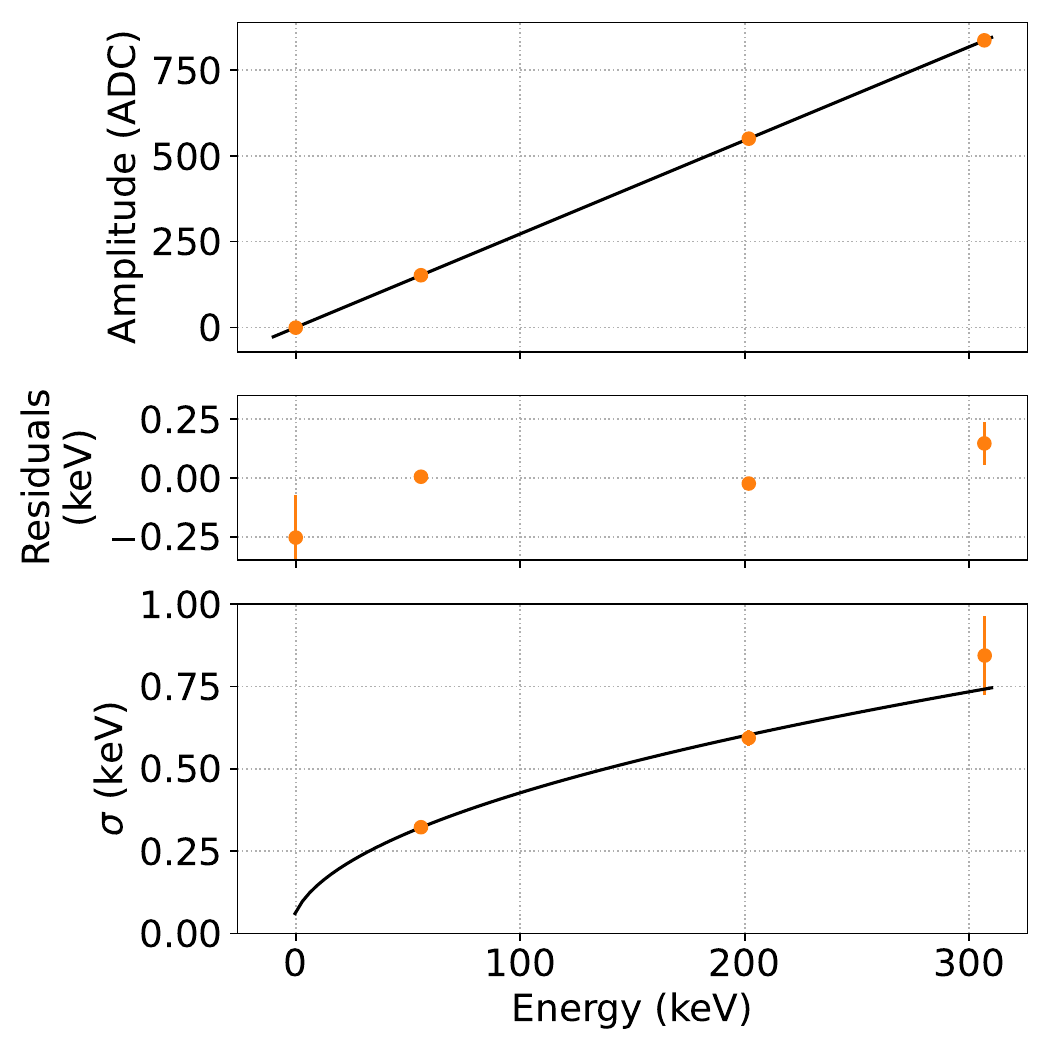}
\caption{Top panel: data points used for the energy calibration, together with the linear fit. Central panel: residuals of the linear fit. Bottom panel: energy resolution $\sigma$ as a function of the photon energy, together with a square-root fit.}
\label{calib}
\end{figure}

Panel \textbf{a} shows all the events measured with the SDD. Panel \textbf{b} (\textbf{c}) shows instead the spectrum of events measured with the SDD in anti-coincidence (coincidence) with the SiPM in blue (orange). In \textbf{b}, events from the beta decay of $^{99}$Tc are clearly visible up to ~300 keV. The Q-value of $^{99}$Tc, as measured by the MetroBeta group \cite{Paulsen2024}, is indeed 295.82(16) keV. Events above 300 keV are background events. In \textbf{c}, the coincidence spectrum is dominated by photon lines associated with the decay of $^{176}$Lu in the LYSO crystal. The peaks at 201 and 306 keV originate from $\gamma$-ray de-excitation of $^{176}$Hf, while the structures around 50–60 keV are mainly due to Hf and Lu X-rays. The underlying continuum is mainly due to Compton scattering of high-energy photons. Since the X-ray region contains several overlapping Hf and Lu contributions, only the most isolated Hf line at 55.8 keV was used for calibration, together with the $\gamma$-ray peaks and the zero-energy peak. The peak positions were extracted with Gaussian fits and used to perform a linear energy calibration. The calibration curve and the corresponding residuals are shown in the first two panels of Figure \ref{calib}. We also tried a quadratic calibration, which produced results compatible with those obtained from the linear fit. We therefore opted for the simpler linear model to describe the SDD energy scale.

\begin{figure}[h]%
\centering
\includegraphics[width=0.5\textwidth]{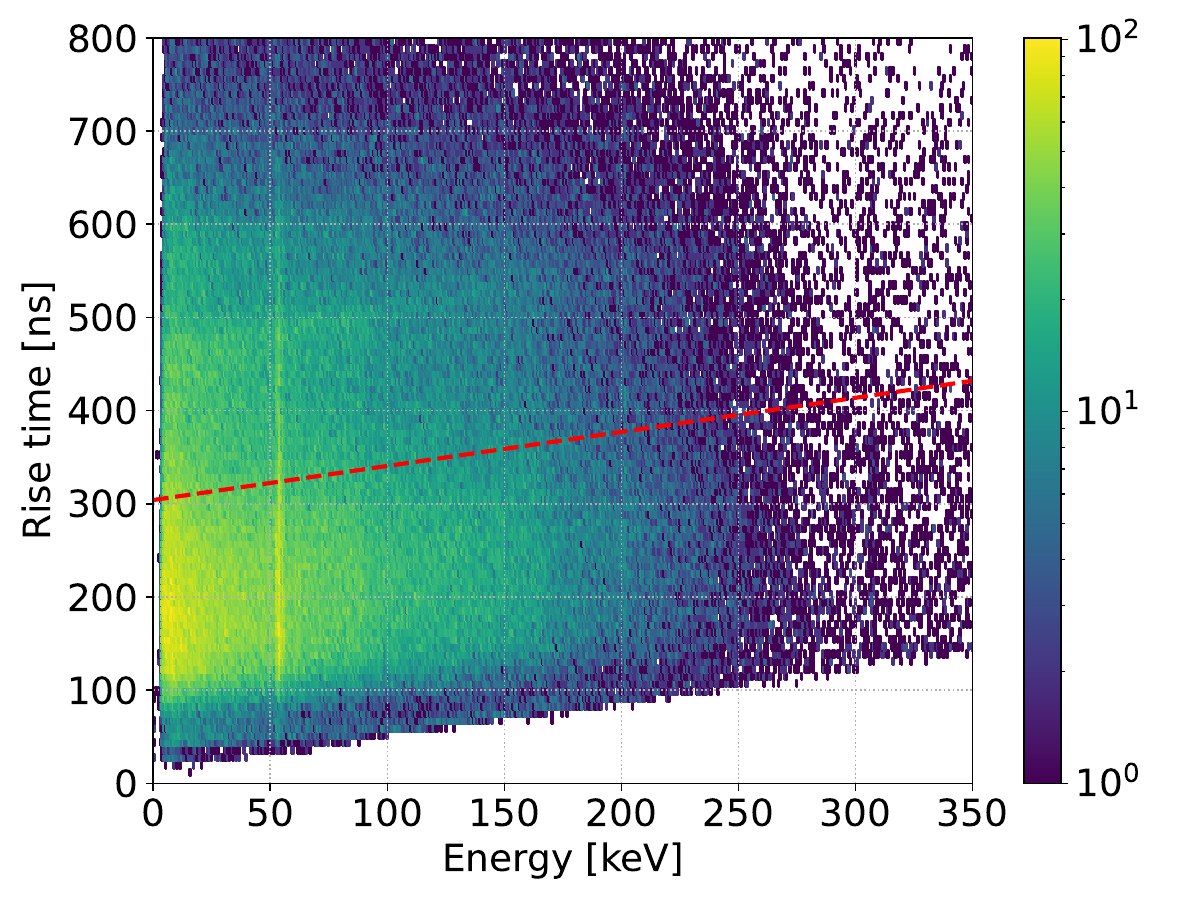}
\caption{2D histogram of rise time and energy for all the events measured by the SDD. The rise time cut used in the analysis is also plotted with a red dashed line.}
\label{rise_ene}
\end{figure}

The residuals of the linear calibration show a maximum deviation below 250 eV over the calibrated energy range. Therefore, we expect our calibration to be accurate within this range. It should be noted that similar precision can be achieved across the entire energy range of the $^{99}$Tc $\beta$ decay thanks to the presence of a $\gamma$ peak in the calibration dataset above the $^{99}$Tc endpoint. \\
Instead, the energy resolution $\sigma$ as a function of photon energy is shown in the bottom panel of Figure \ref{calib}. The point corresponding to the zero-energy peak had an uncertainty that was too large, so it was not included in the plot. A square-root fit of the data points confirms the expected behaviour of the energy resolution. \\
Although the large active area of the SDD reduces edge effects, charge loss still has to be considered for an accurate spectral analysis, since it shifts events toward lower reconstructed energies and modifies the spectral shape. One way to eliminate this systematic effect is to reject events based on their rise time. In a Silicon Drift Detector, the electrons produced after ionising radiation interaction are collected from a central anode thanks to radial drift. The rise time of the signal is proportional to the width of the electron cloud at the anode. This width is larger for events with an interaction point close to the detector border because the width of the cloud increases due to thermal diffusion while drifting towards the anode. Therefore, keeping only events with lower rise times helps to reduce the charge loss systematics. However, this is complicated by the fact that the width of the electron cloud increases with the energy-dependent distance travelled by the ionising electron, which is larger for higher energy particles. Thus, the minimum rise time increases with electron energy. Figure \ref{rise_ene} shows the distribution of rise time computed as the distance between the 10\% and 90\% amplitude points as a function of energy, where this effect can be clearly seen. 

\begin{figure}[h]%
\centering
\includegraphics[width=0.5\textwidth]{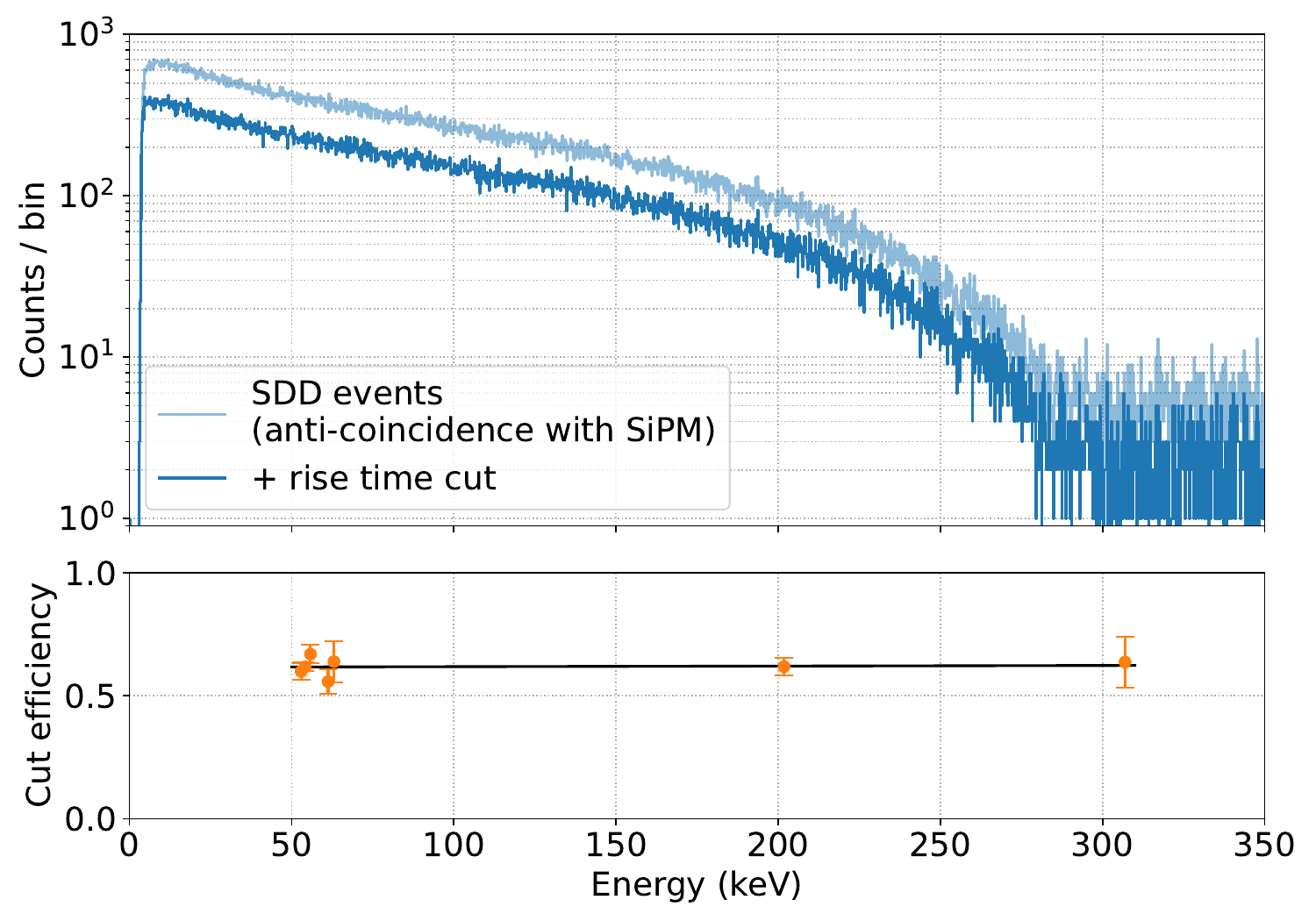}
\caption{Top panel: anti-coincidence spectrum before and after the rise time cut. Bottom panel: efficiency of the rise time cut evaluated from the X and $\gamma$ ray peaks of the coincidence spectrum.}
\label{cut}
\end{figure}

We adopted a linear rise-time cut, fixing its slope to correspond to the minimum rise time as a function of energy. We then selected an offset value as the smallest that ensured the cut of edge-like events while maintaining a stable spectral shape. Using a lower offset value would also cut non-edge-like events. The rise time cut line is shown in red dashed lines in Figure \ref{rise_ene}. \\
The SDD spectrum in anti-coincidence with the SiPM before and after the rise time cut is shown in Figure \ref{cut}. An important quantity to estimate after applying the cut is its efficiency. In particular, it is important to determine whether there is any energy dependence, as this would lead to a distortion that must be taken into account in spectral analysis. The cut efficiency was evaluated by computing the ratio of the areas of the X-ray and gamma-ray peaks in the calibration spectrum before and after the rise-time cut. The efficiency of each data point is shown in the bottom panel of Figure \ref{cut}. We performed a linear fit of these points and found a slope compatible with zero. Therefore, we conclude that no energy-dependent efficiency needs to be considered after the rise time cut. We also found that changing the linear rise-time cut offset slightly produced a spectrum with a compatible shape and flat cut efficiency. In contrast, changing the rise-time cut slope resulted in an energy-dependent cut efficiency. Therefore, we can conclude that our choice of rise-time cut parameters is robust and does not introduce a significant systematic uncertainty into the analysis. A total of approximately 180 thousand events survived the cut and were adopted in the analysis.

\begin{figure}[h]%
\centering
\includegraphics[width=0.5\textwidth]{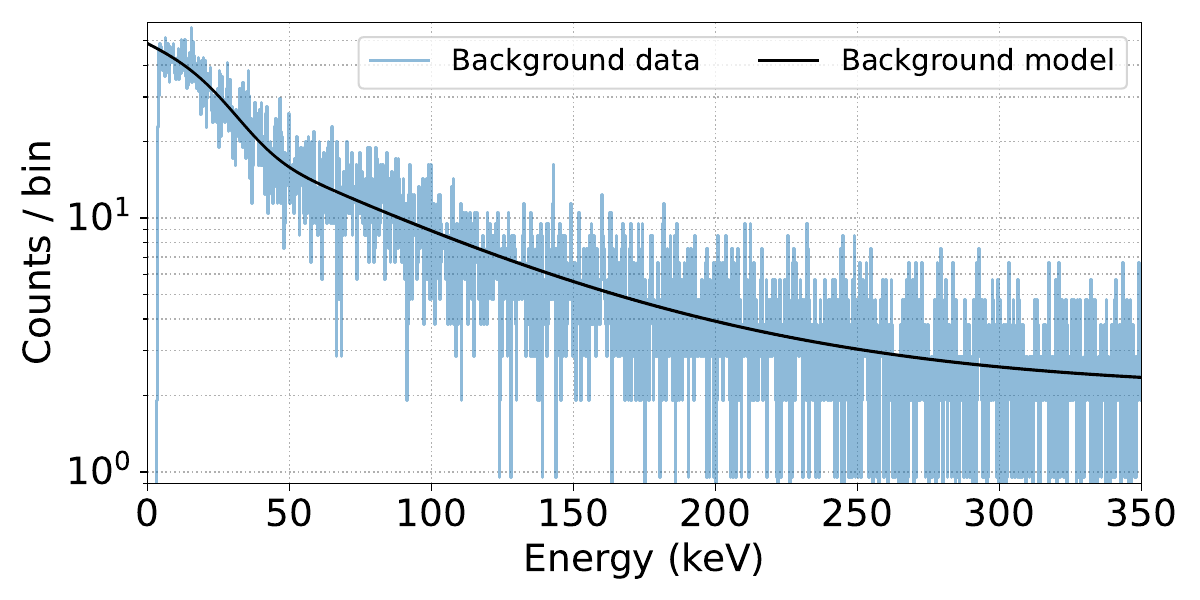}
\caption{Background measured over six days by substituting the silicon substrate with the Tc deposition with a bare one (blue). An analytical model fitted on background data is also shown (black).}
\label{bkg}
\end{figure}

As discussed above, background events also play a role in our measured spectrum. To be able to compare the $^{99}$Tc spectrum with theoretical predictions, these have to be included in the analysis. Another advantage of our setup is that we can measure the background spectrum directly and include it in the analysis chain. This was achieved by replacing the silicon substrate carrying the Tc deposition with a bare silicon substrate. We then performed another six-day measurement under the same experimental conditions. We used the photon peaks from the LYSO crystal to perform a linear calibration and applied the same rise time cut. This verified that the efficiency was not energy-dependent in this case either. The resulting spectrum, which contains background events only, is shown in Figure \ref{bkg}. It is worth noting that background events can originate from radioactive decay in components close to our setup or from the LYSO crystal in cases where insufficient light is collected on the SiPM to detect a pulse above the threshold. Directly measuring the background in the same configuration as the $^{99}$Tc measurement ensures that all of these possibilities are taken into account. The measured background is featureless; therefore, to eliminate statistical fluctuations, we developed an analytical background model consisting of a low-energy error function, a higher-energy exponential and a constant component. The parameters of this model were fitted to the measured background. The best fit is shown in black in Figure \ref{bkg}. In the following, this background model is used when fitting the $^{99}$Tc data.

\section{\label{sec:met} Analysis method}

\begin{figure}[h]%
\centering
\includegraphics[width=0.5\textwidth]{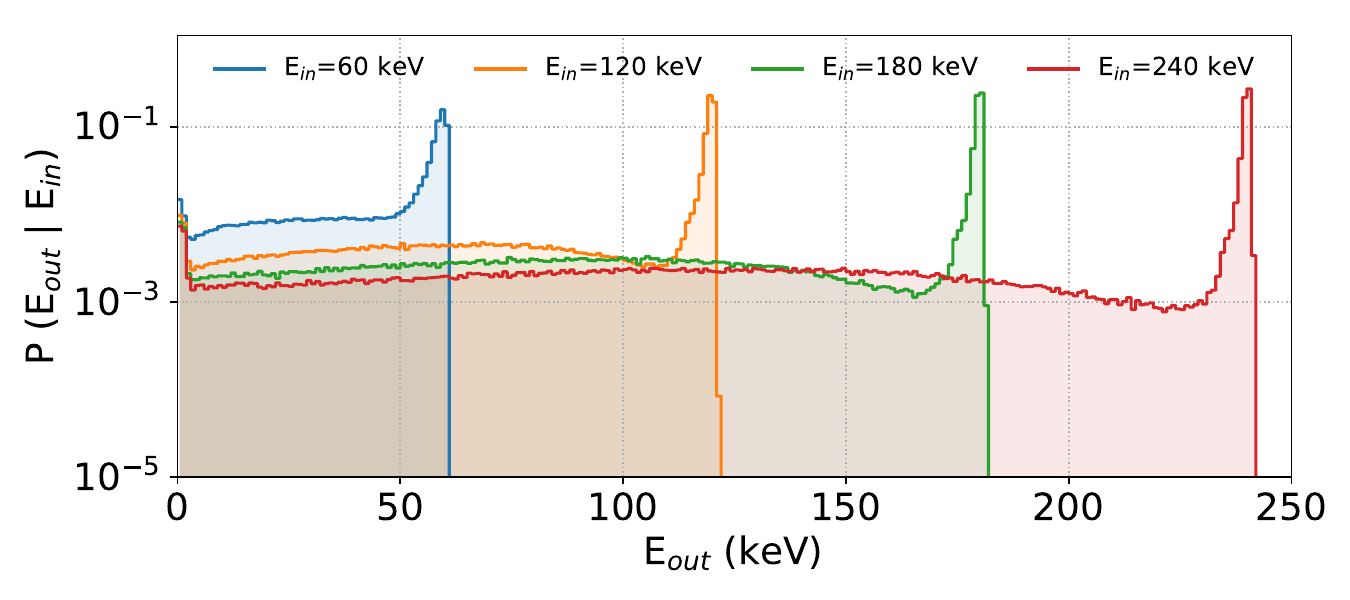}
\caption{SDD experimental response for different electron energies simulated with GEANT4. The detector response has been added to the GEANT4 output. The asymmetric peak due to autoabsorption as well as the low-energy tail due to detector backscattering can be clearly seen.}
\label{resp}
\end{figure}

In this section, we discuss the method used to analyze the measured beta spectrum. As discussed in Section \ref{sec:intro}, effects such as autoabsorption in the Technetium source, as well as backscattering on the detector, cause a distortion of the spectrum shape, which has to be taken into account by means of Monte Carlo simulations. \\
We developed a GEANT4 \cite{Agostinelli2003} simulation of the whole setup, exploiting the physics package Penelope \cite{Asai2021}, which is optimized for low-energy electromagnetic interactions. Concerning the SDD, the effect of a dead layer is taken into account as shown in \cite{Nava:2024wsa}, while the energy resolution is added through a convolution with an energy-dependent Gaussian, whose width increases with the square root of the energy. The energy resolution $\sigma$ scaling was fixed to the one extracted from calibration data (shown in Figure \ref{calib}). The Tc source has instead been simulated as a uniform layer on top of a silicon substrate. Electrons were generated in this technetium volume with an isotropic direction. In practice, the deposited Tc could produce small crystals on the silicon surface, with a thickness that may vary from one crystal to another. The Tc thickness that we use in our simulations is therefore an effective thickness parametrizing the source autoabsorption. \\
Simulations have been run for monochromatic electrons of different energies E$_{in}$ produced in the source and absorbed by the detector. The distribution of the energy measured by the detector E$_{out}$, normalized to one, is hereby indicated with P(E$_{out}$$|$E$_{in}$). This response, including detector effects, simulated for a Tc thickness of 500 nm, is reported in Figure \ref{resp} for four different E$_{in}$.

\begin{figure}[h]%
\centering
\includegraphics[width=0.5\textwidth]{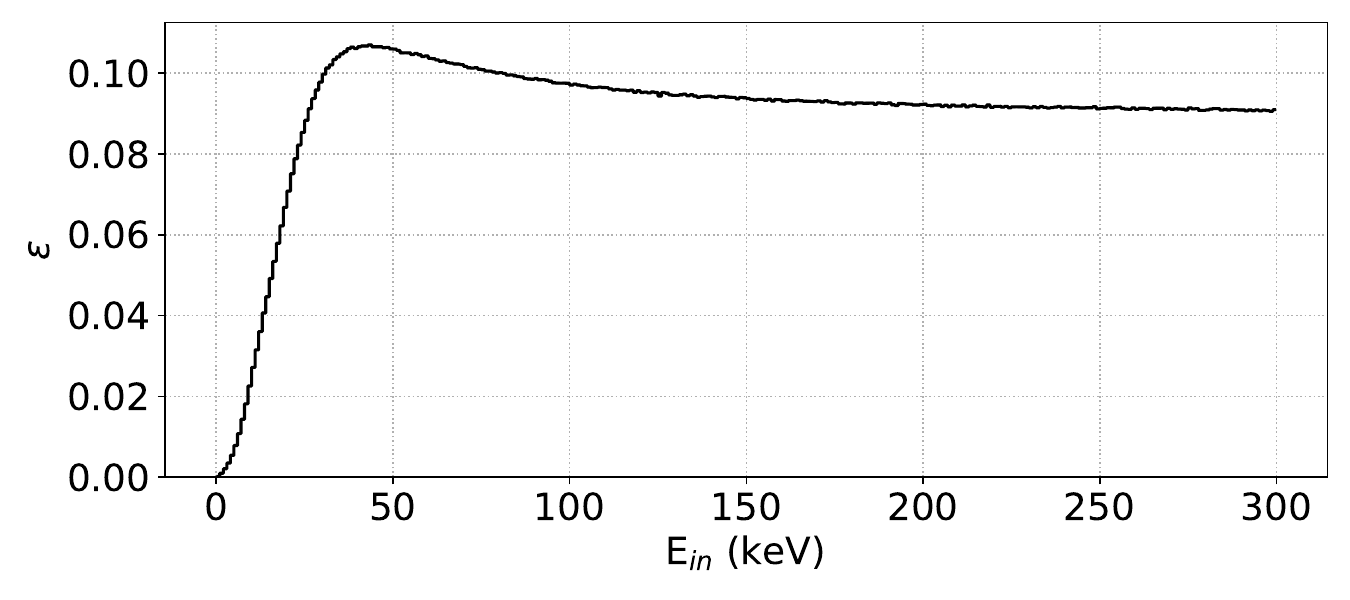}
\caption{Efficiency $\epsilon$ as a function of the electron energy simulated with GEANT4. The quantity $\epsilon$ is defined as the fraction of the simulated electrons that deposit energy in the active volume of the SDD. }
\label{eps}
\end{figure}

It can be seen how the response to monochromatic electrons is non-Gaussian, as expected. An asymmetric peak is present close to E$_{in}$, with a Gaussian right side dominated by energy resolution and a left side dominated by energy loss in the source before hitting the detector. A continuous tail can also be seen, due to backscattering at the detector. It goes down to zero, since electrons have the chance to leave the detector without depositing energy in the active volume. It must also be noted that the experimental response is energy-dependent. In particular, by looking at the left side of the peak, the influence of source autoabsorption is more prominent for electrons with lower E$_{in}$. \\
As already stated, the experimental responses have been normalized, but the efficiency $\epsilon$, which is the fraction of the generated electrons that actually deposit energy in the SDD, for a given E$_{in}$, can also be obtained from simulations. This quantity, computed from the same GEANT4 simulation already mentioned, is shown in Figure \ref{eps}.

\begin{figure}[h]%
\centering
\includegraphics[width=0.5\textwidth]{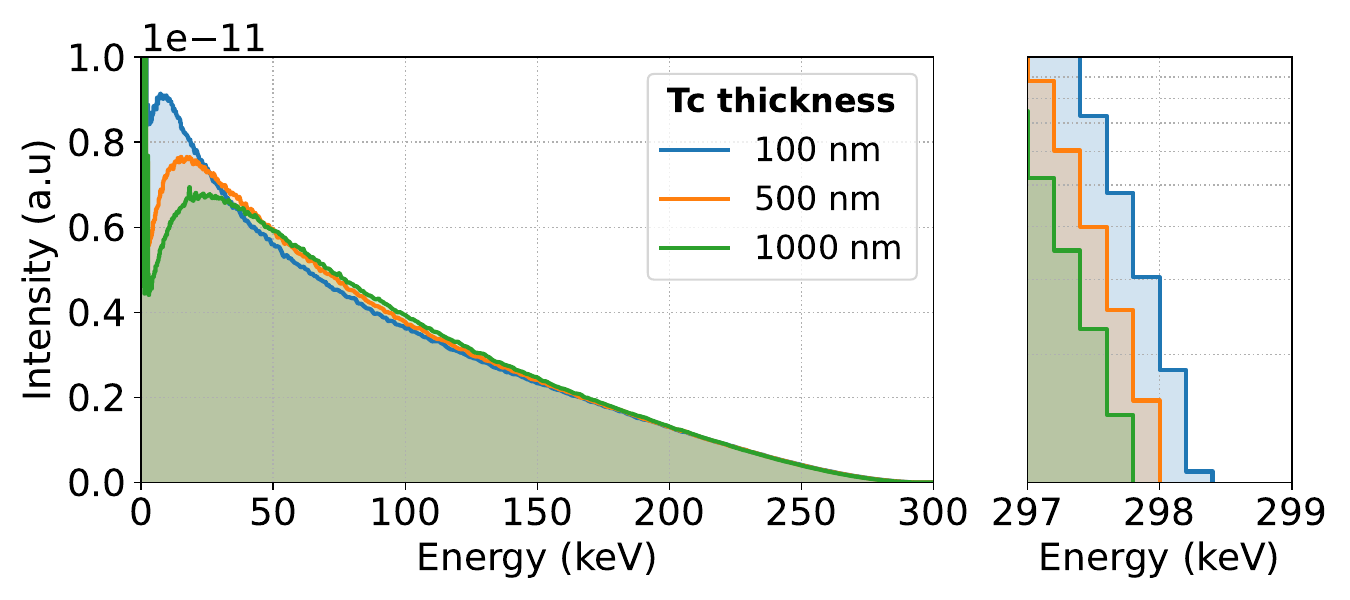}
\caption{Predicted measured $^{99}$Tc spectrum for three different Tc thicknesses in the GEANT4 simulation. The left panel shows the full spectrum, while the right panel shows a zoom of the endpoint region.}
\label{model}
\end{figure}

It can be seen that the efficiency for E$_{in}$ $>$ 0 is always non-zero, due to the fact that the Tc source has not been encapsulated between passive materials. However, a low-energy cutoff effect is visible, still due to source autoabsorption. For higher energy, instead, the efficiency asymptotically converges to the geometric efficiency of $\sim$9\% of our setup. This means that the effect of autoabsorption in the source is mostly important for the low-energy region of a $\beta$ spectrum. \\
Denoting by T(E$_{in}$) a theoretical prediction of the $^{99}$Tc $\beta$ spectrum, the spectrum measured by the SDD, M(E$_{out}$), can be computed with the following formula:
\begin{equation}
\label{eq:model}
    M(E_{out}) = \int_0^Q P(E_{out}|E_{in}) \cdot \left(\epsilon(E_{in}) \cdot T(E_{in})\right)dE_{in} + B
\end{equation}
where $B(E_{out})$ is the background model described in the previous section. For binned objects, this formula translates into a matrix multiplication. P(E$_{out}$$|$E$_{in}$) is called the response matrix, and its columns are given by response functions such as those shown in Figure \ref{resp}. The response matrix depends on one free parameter that is not constrained by any calibration measurement in our setup, namely the Tc thickness. Figure \ref{model} shows the predicted measured spectrum M(E$_{out}$) for three different Tc thicknesses.

As expected, the largest effect due to changing the Tc thickness can be seen at low energies. A thicker (thinner) Tc layer produces a cutoff at higher (lower) energies. The right panel shows instead a zoom of the endpoint region, in logarithmic scale, where a small shift of the measured endpoint can be seen as well. \\
In order to fit the measured spectrum, we performed a scan over the Tc thickness $t$ of the following $\chi^2$:
\begin{equation}
    \chi^2 (t) = \sum_{i=1}^{N_{bins}}\left(\frac{D(E_{out})^i - M(E_{out}| t)^i}{\sigma_D(E_{out})^i}\right)^2
\end{equation}
where D(E$_{out}$) is the array containing the experimental spectrum and $\sigma_D$(E$_{out}$) is the related Poissonian uncertainty. The fit has been performed in the energy range between 5 and 350 keV. The spectrum prediction is normalized to the data in this energy range; therefore, only the information on the shape of the spectrum is used in the fit.

\section{\label{sec:res} Results and discussion}
\begin{figure}[h]%
\centering
\includegraphics[width=0.5\textwidth]{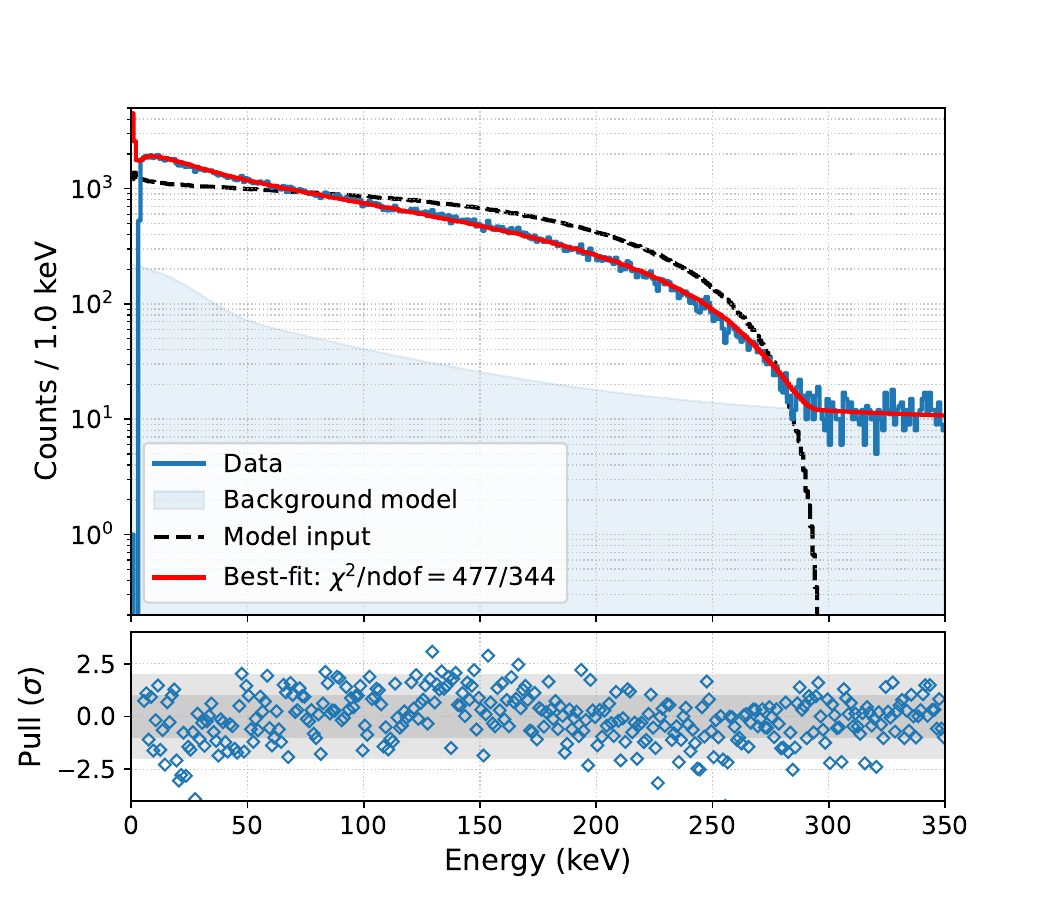}
\caption{Top panel: comparison between our SDD measurement (blue) and the MetroBeta measurement \cite{Paulsen2024} processed with our Monte Carlo-based method to include the experimental response (red). The fit is performed with only one free parameter, being the Tc thickness. The MetroBeta spectrum used as an input is shown in black. Bottom panel: residuals of the fit.}
\label{first_fit}
\end{figure}

\textit{Comparison with other data} - Before comparing with nuclear-model predictions, we first assess whether our measured spectrum is compatible with previous $^{99}$Tc measurements. In particular, we performed a fit using the MetroBeta spectrum \cite{Paulsen2024} as the input T(E$_{in}$) to our response model. The $\chi^2$ minimization yielded a best-fit value of $\sim$180 nm for the Tc thickness. The resulting best-fit spectrum is shown in red in Figure \ref{first_fit}, together with the SDD data in blue. The input spectrum, measured with MMCs, is shown in black, while the background model is shown with a shaded blue area. The bottom panel reports instead the fit residuals.

It can be seen how our spectrum looks very different with respect to the MetroBeta one, measured with MMCs. The difference is due to the systematic effects related to having the $^{99}$Tc source not embedded in the detector active volume, mainly electron backscattering. However, it can be noted how, after folding the MetroBeta spectrum with our simulated experimental response and adding our measured background, the agreement improves significantly. In particular, both the low-energy region of the spectrum and the background-dominated region are very well reproduced. The p-value of the fit, however, indicates a statistically poor reproduction of the spectrum. It must be noted that the MetroBeta spectrum used as an input is not directly measured; detector effects have been unfolded, and background has been subtracted, making some assumptions on its shape. It is therefore possible that some systematic effect, in their model or in our model, is not accurately taken into account, although we performed our simulation to the best of our knowledge of the experimental setup.

\begin{figure}[h]%
\centering
\includegraphics[width=0.5\textwidth]{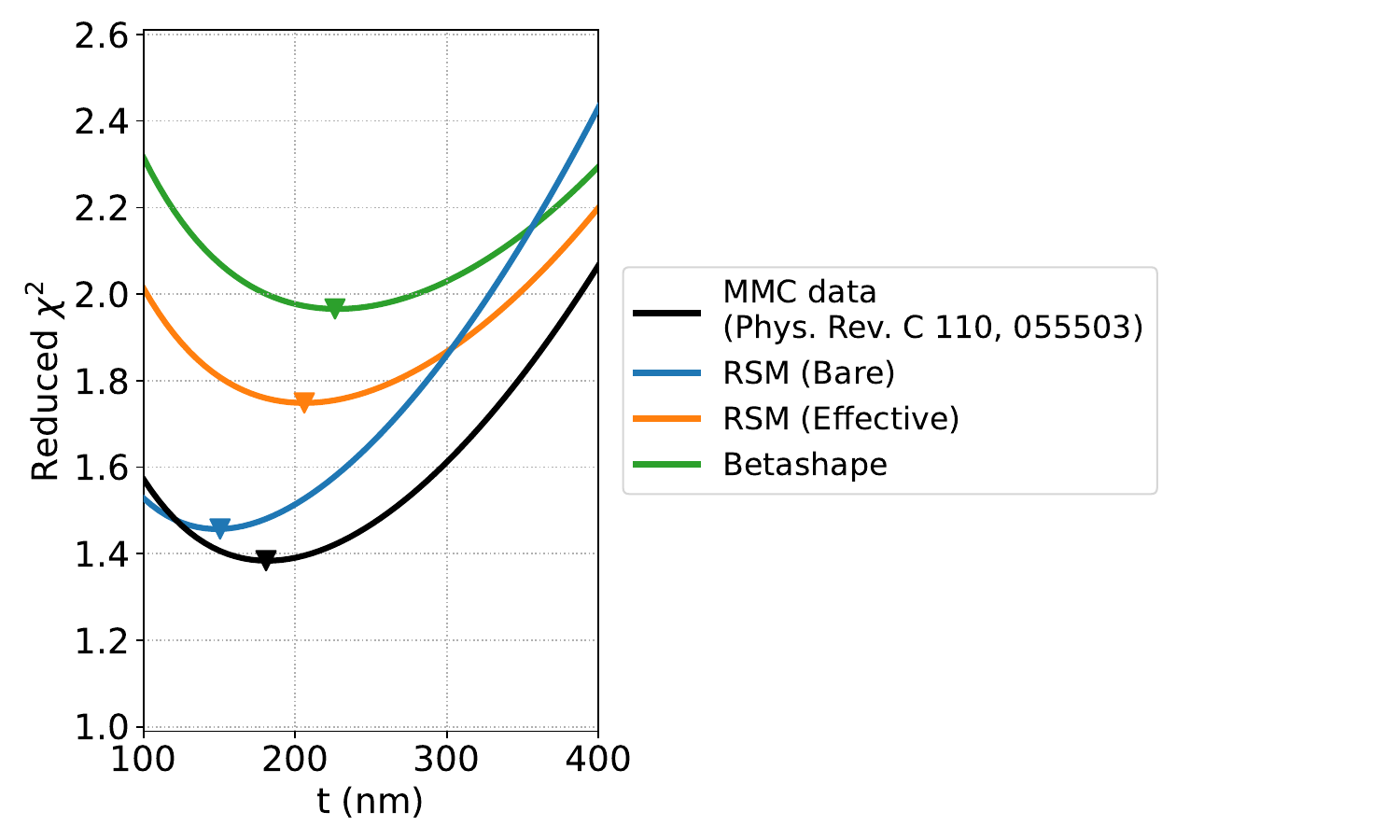}
\caption{Reduced $\chi^2$ as a function of the Tc thickness $t$ computed using as an input the MetroBeta data and three different theoretical spectra: Bare and Effective from the RSM, and the one from the Betashape model. The coloured triangles represent the absolute minima of these curves.}
\label{chi2}
\end{figure}

\textit{Comparison with nuclear models} - 
We compared our measured spectrum with different nuclear-model predictions. We considered the spectrum calculated with \textsc{BetaShape} \cite{Mougeot2015} and those obtained within the Realistic Shell Model (RSM) \cite{PhysRevC.110.014324}, in which the Hamiltonian and effective decay operators were derived microscopically from the CD-Bonn nucleon--nucleon potential \cite{Machleidt2001} using the $V_{\mathrm{low}\text{-}k}$ procedure \cite{Bogner2002}. 
In particular, we considered two RSM  spectra, one obtained with unrenormalized decay operators (Bare), and one using decay operators derived consistently with the effective Hamiltonian (Effective), the relativistic nuclear matrix element is fixed according to the CVC hypothesis \cite{BehrensBuhring1971}. It is worth mentioning that the same effective Hamiltonian was previously employed in the RSM study of the double-$\beta$ decay of $^{100}\mathrm{Mo}$ \cite{Coraggio2022}. At this stage, the fit was based only on the measured spectral shape, without imposing a constraint from the experimental half-life. For each input spectrum, we varied the Tc thickness and folded the theoretical prediction with the corresponding detector-response model. The resulting reduced $\chi^2$ profiles are shown in Figure~\ref{chi2}.

None of the nuclear-model predictions reproduces the measured spectrum as well as the MetroBeta input. Nevertheless, both RSM calculations provide a better description of the spectral shape than \textsc{BetaShape}, supporting the microscopic approach adopted in the RSM. Interestingly, when only the spectral shape is considered, the Bare calculation performs better than the Effective one. Conversely, the Effective calculation provides a more accurate prediction of the experimental half-life \cite{PhysRevC.110.014324}. This suggests that the spectral shape and half-life provide complementary constraints and motivates their simultaneous inclusion in the following analysis.


\textit{Fit with free nuclear model parameters} - 
On the above basis, following the approach adopted in the MetroBeta analysis \cite{Paulsen2024}, we performed a combined fit of the spectral shape and half-life starting from the RSM calculation with Bare decay operators and treating $g_A$ and $g_V$ as free parameters. Their fitted values are reported relative to the reference couplings $g_A^{\mathrm{ref}}=1.2723$ and $g_V^{\mathrm{ref}}=1$ through the factors $q_{g_A}=g_A/g_A^{\mathrm{ref}}$ and $q_{g_V}=g_V/g_V^{\mathrm{ref}}$. A single value is used for all contributions associated with each current, so that $q_{g_A}$ and $q_{g_V}$ can be interpreted as effective average quenching factors for the axial and vector currents, respectively. For each pair $(q_{g_A},q_{g_V})$, the RSM predicts both the spectral shape and the half-life, which are simultaneously compared with the measured spectrum and the experimental half-life, $T_{1/2}=2.111(12)\times10^{5}$ y \cite{BROWNE201725}.

\begin{figure}[h]%
\centering
\includegraphics[width=0.5\textwidth]{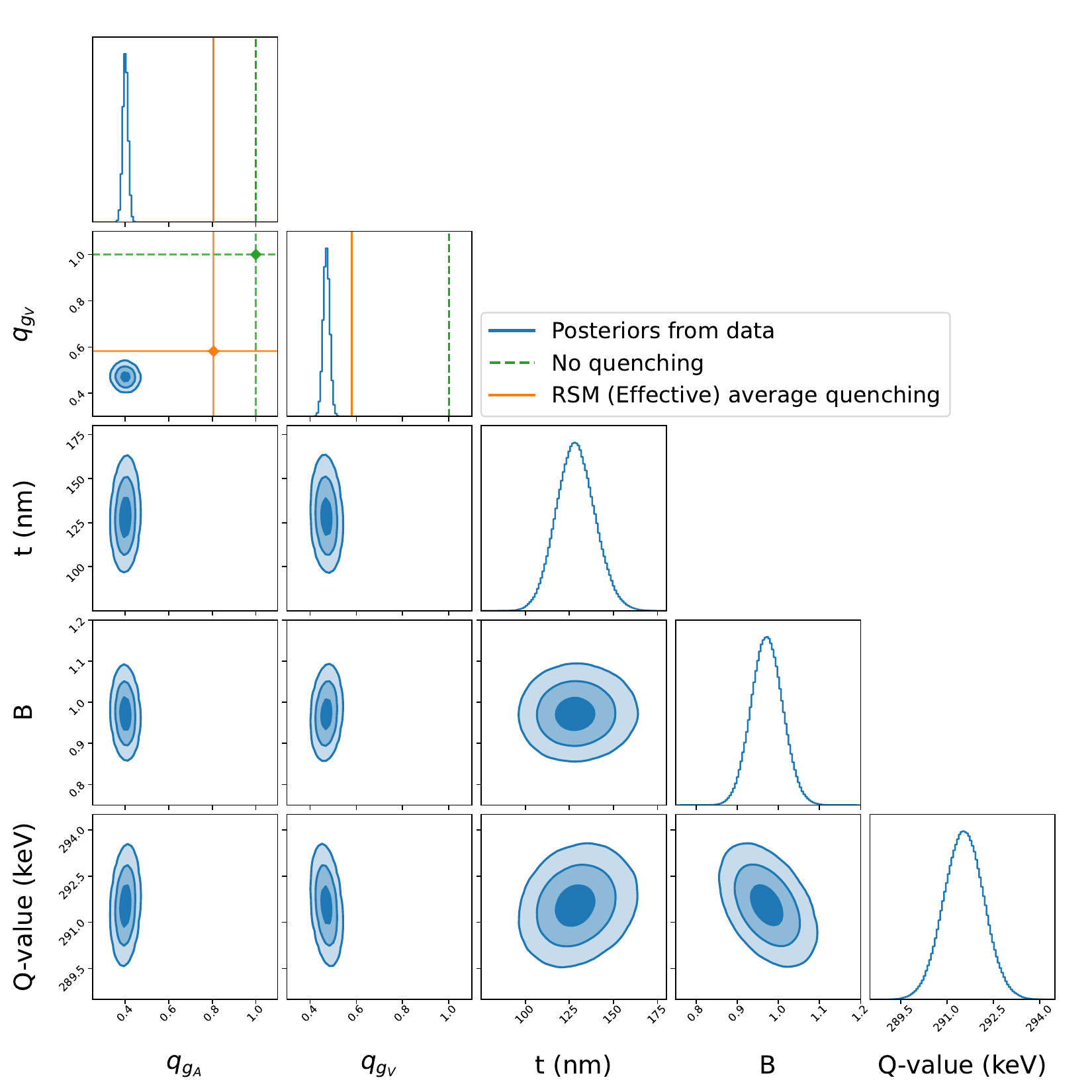}
\caption{Corner plot of the joint and marginalized posteriors of the 5 parameters left free in our model. An uninformative flat prior has been used for all the parameters. The result starting from the Bare RSM is shown in blue. The green lines indicate the no-quenching hypothesis, while the orange lines indicate the effective average quenching obtained from the Effective RSM.}
\label{corner}
\end{figure}

To ensure the robustness of our analysis, we have left the thickness of the Tc $t$ and the background amplitude $B$ free in the fit, parametrising the imperfect knowledge of the main systematics in our experimental setup. Finally, the Q-value of the decay has also been included as a free parameter, as it contributes to determining both the spectral shape and the half-life. Therefore, this analysis can also be used to determine the Q-value from our data, which cannot be done directly from the measured spectrum. If effects such as source autoabsorption are not taken into account, they can cause a bias in the recovery of the Q-value (as shown in Figure \ref{model}).
Due to the complexity of this model, to ensure the convergence of the fit and a robust recovery of the parameters, we performed a Bayesian sampling of the log-likelihood function, defined as

\begin{equation}
\label{eq:like}
\begin{split}
log\mathcal{L}(p) = -\frac{1}{2}\Bigg[
&\sum_{i=1}^{N_{\mathrm{bins}}}
\left(
\frac{D^i - M(p)^i}{\sigma_D^i}
\right)^2 \\
&+
\left(
\frac{T_{1/2}^D - T_{1/2}^M(p)}
{\sigma_{T_{1/2}^D}}
\right)^2
\Bigg],
\end{split}
\end{equation}

where $p$ indicates the parameter array, $q_{g_A}, q_{g_V}, t, B, Q$. The first term of the likelihood is a $\chi^2$ statistic computed over the spectrum bins, as in the previous section. Therefore, it represents the penalty for not reproducing the spectral shape accurately. The second term instead represents a Gaussian penalty for not reproducing the half-life $T_{1/2}$ accurately. Note that, referring back to equation \ref{eq:model}, the parameters $(q_{g_A}, q_{g_V}, Q)$ are used in the calculation of T(E$_{in}$). For a given set of these parameters, both the theoretical spectral shape and the half-life are computed. $t$ enters into the computation of the response matrix, while $B$ is simply a scaling factor of the background model that already appears in that equation. We chose to use flat, uninformative priors on all the parameters so that sampling the Bayesian posterior essentially coincides with sampling the likelihood in equation \ref{eq:like}. \\
The half-life component of the likelihood can be considered a multidimensional prior in the five-dimensional parameter space being sampled, favouring only combinations of parameters that reproduce the experimental half-life. The posterior sampling was performed using the Python package \texttt{emcee} \cite{Foreman_Mackey_2013}. The resulting posteriors, obtained from the Bare RSM calculations, are shown in the corner plot in Figure \ref{corner}. The best-fit parameters were then extracted by computing the posterior maximum. Table \ref{tab:fit} summarises the best-fit parameters, together with the values obtained by MetroBeta and those computed using the effective RSM.

\begin{table}[h]
\renewcommand{\arraystretch}{1.3} 
\resizebox{0.5\textwidth}{!}{%
\begin{tabular}{c|ccc}
\hline
Parameter & Best-fit & RSM (Effective) & MetroBeta \cite{Paulsen2024} \\
\hline
$q_{g_A}$ &  0.40 $\pm$ 0.01 & 0.81 & 0.452 $\pm$ 0.028 \\
$q_{g_V}$ & 0.47 $\pm$ 0.01 & 0.58 & 0.376 $\pm$ 0.005 \\
t (nm) & 128 $\pm$ 11 & &  \\
B & 0.97 $\pm$ 0.04 & &  \\
Q-value (keV) & 291.6 $\pm$ 0.7 &  & 295.82 $\pm$ 0.16 \\
\hline
\end{tabular}
}
\caption{Best-fit parameters obtained by sampling with an MCMC the likelihood in equation \ref{eq:like}. The average quenching computed from the Effective RSM, as well as the values obtained by MetroBeta, are reported as well.}
\label{tab:fit}
\end{table}

The fit yields $q_{g_A}=0.40(1)$ and $q_{g_V}=0.47(1)$, indicating a quenching of both the axial and vector currents. MetroBeta obtained $q_{g_A}=0.452(28)$ and $q_{g_V}=0.376(5)$. The two determinations of $q_{g_A}$ agree within approximately two standard deviations, whereas those of $q_{g_V}$ are not statistically compatible. However, a direct agreement is not necessarily expected because the two analyses rely on different effective Hamiltonians and, consequently, different nuclear wave functions.


We then compared the experimentally inferred quenching factors with the average renormalization predicted by the RSM calculation employing microscopically-derived effective decay operators. This renormalization was quantified by taking the ratios of the eight NMEs calculated with the Effective and Bare operators: four associated with axial operators and four with vector operators \cite{PhysRevC.110.014324}. The corresponding mean ratios give $q_{g_A}=0.81$ and $q_{g_V}=0.58$. These values are reported in Table~\ref{tab:fit} and indicated by the orange lines in Figure~\ref{corner}, while the green lines represent the no-quenching hypothesis. For both currents, the microscopic renormalization shifts the RSM predictions towards the region favored by the fit. The effective value of $q_{g_V}$ lies relatively close to the fitted posterior, whereas the renormalization of the axial current moves in the same direction but remains weaker than the quenching required by the data. Thus, the effective RSM operators reproduce the overall trend inferred from the experiment, although they do not fully account for the magnitude of the extracted quenching.

Examining the other fit parameters reveals that the experimental ones are well constrained. This suggests that an imperfect understanding of the response of the setup does not prevent the precise estimation of nuclear model parameters. In our specific case, this is possible because the thickness of the Tc is constrained by the low-energy region of the spectrum and the background amplitude from the region above the Q-value. The latter is compatible with 1 within one sigma, meaning that the background amplitude used as an input, which came directly from an experimental measurement, was accurate. This is possible thanks to the flexibility of our setup when performing measurements with and without the radioactive source. 

\begin{figure}[h]%
\centering
\includegraphics[width=0.5\textwidth]{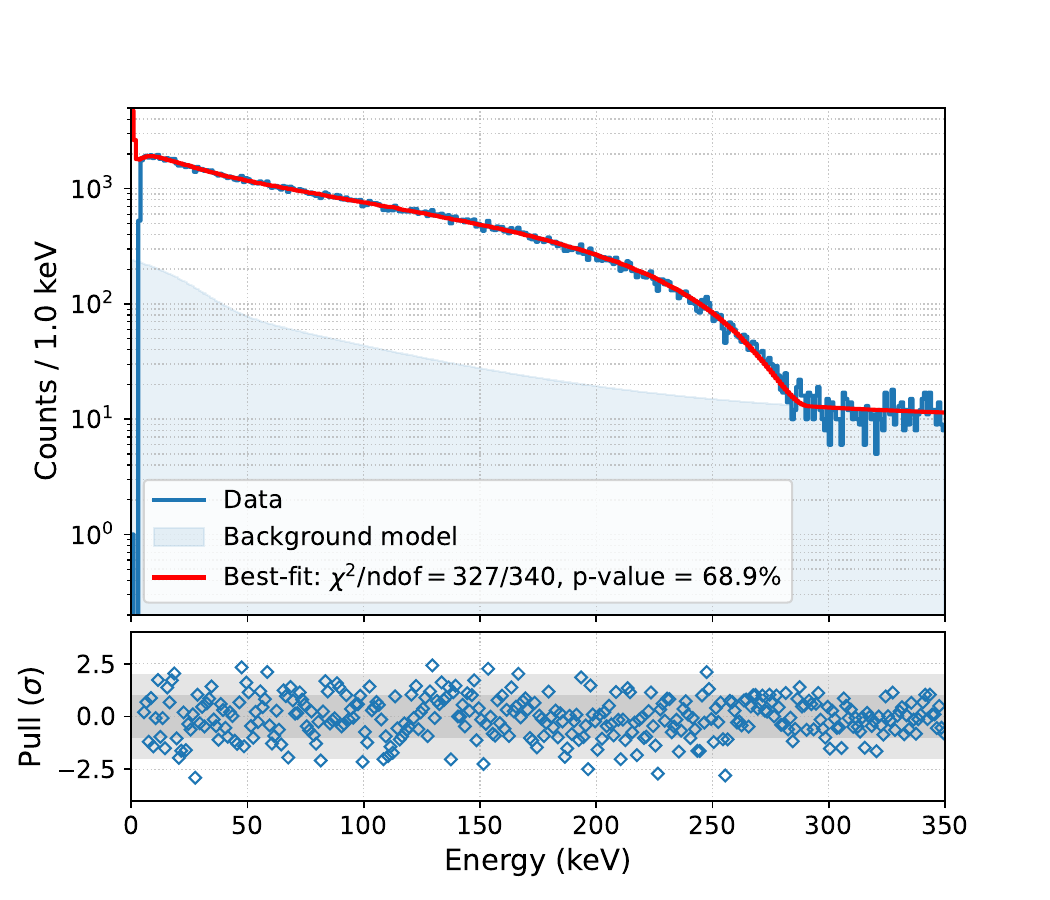}
\caption{Top panel: comparison between our SDD measurement (blue) and the best-fit spectrum obtained from the Enhanced Spectral Shape Method, starting from the Bare RSM prediction (red). The Bayesian fit is performed with five free parameters. Bottom panel: residuals of the fit.}
\label{best_fit}
\end{figure}

Finally, we estimated a Q-value of 291.6(7) keV. This value is lower than, and incompatible with, that measured by the MetroBeta group (295.82(16) keV). The best-fit result reported in Table \ref{tab:fit} only considers statistical uncertainty. We expect the main systematic contribution to come from the energy calibration. However, we estimated a systematic contribution of 0.25 keV to the Q-value determination from the fit result in Figure \ref{calib}. Even when this systematic uncertainty is taken into account, the discrepancy remains. To evaluate the impact of this discrepancy on the estimation of nuclear parameters, we performed a second fit using a Gaussian prior on the Q-value whose position and width were derived from the MetroBeta measurement. The values of $q_{g_A}$ and $q_{g_V}$ obtained in this way are 0.42(1) and 0.43(1), respectively. These values are similar to those previously reported and do not change any of our conclusions regarding the renormalisation of the axial and vector currents. 
The best-fit spectrum computed from the parameters in Table \ref{tab:fit} is shown in Figure \ref{best_fit}, compared to SDD data. 

The computed p-value of approximately 70\% now indicates statistical agreement with the experimental data. Although a shape is still visible in the residuals, the quality of the reconstruction is noticeably better than that shown in Figure \ref{first_fit}, particularly at low energy and in the Q-value region. \\

\section{\label{sec:disc} Conclusion}
The second-forbidden non-unique $\beta$ spectrum of $^{99}$Tc was measured for the first time with a Silicon Drift Detector down to 5 keV within the ASPECT-BET project. We developed a hybrid spectrometer combining the SDD with a LYSO scintillator read by a SiPM. Coincidence events allowed us to calibrate the SDD during the $^{99}$Tc measurement without subtracting the calibration-source contribution from the $\beta$ spectrum. The background was measured separately under the same experimental conditions, while the rise-time selection was verified to have no significant energy dependence. We also developed a detailed GEANT4 simulation of the experimental response, including source self-absorption, electron backscattering, detection efficiency, and energy resolution. After including the simulated response of our setup, the MetroBeta spectrum \cite{Paulsen2024} showed decent agreement with the SDD data, indicating that the differences between the directly measured spectra are mainly due to experimental effects. The measured spectrum was also compared with the spectral shapes predicted by \textsc{BetaShape} and the RSM using Bare and Effective operators. Both RSM calculations reproduced the measured shape better than \textsc{BetaShape}, with the Bare calculation performing better than the Effective one when only the spectral shape was considered. 
Finally, starting from the Bare RSM operators, we performed a Bayesian fit combining the measured spectrum with the experimental $^{99}$Tc half-life. The fit yielded average quenching factors of $q_{g_A}=0.40(1)$ and $q_{g_V}=0.47(1)$. The renormalization predicted by the Effective RSM moves both currents towards the values preferred by the fit, although it does not fully reproduce the extracted quenching. The fit also yielded a Q-value of $291.6(7)$ keV, which is incompatible with the MetroBeta result. However, repeating the analysis using the MetroBeta Q-value as a Gaussian prior gave similar quenching factors, showing that this discrepancy does not affect our conclusions. 
Overall, this work demonstrates a flexible experimental method for forbidden $\beta$ spectroscopy based on semiconductor detectors, coincidence calibration, direct background measurement, and detailed response simulations. The ASPECT-BET collaboration also plans to measure other forbidden $\beta$ spectra, including those of $^{36}$Cl and $^{94}$Nb, to provide further experimental data for testing nuclear models and the renormalization of decay operators.

\section{\label{sec:ack} Acknowledgements}
We acknowledge ﬁnancial support under the National Recovery and Resilience Plan (NRRP), Mission 4, Component 2, Investment 1.1, Call for tender No. 104 published on 2.2.2022 by the Italian Ministry of University and Research (MUR), funded by the European Union – NextGenerationEU– Project Title ASPECT-BET: An sdd-SPECTrometer for BETa decay studies – CUP H53D23001020006 - Grant Assignment Decree No. 974 adopted on June 30, 2023 by the Italian Ministry of University and Research (MUR).

\bibliography{main}

\end{document}